\documentclass[twocolumn, aps, prd]{revtex4}
\usepackage{graphicx}
\usepackage{amssymb}

\newcommand{\Vec}[1]{\mbox{\boldmath$#1$}}
\def\partd#1#2{\frac{\partial #1}{\partial #2}}

\begin{document}

\title{Hall plateau diagram for the Hofstadter butterfly energy spectrum}
\author{Mikito Koshino and Tsuneya Ando}
\affiliation{
Department of Physics, Tokyo Institute of Technology,
2-12-1 Ookayama, Meguro-ku, Tokyo 152-8551, Japan}
\date{\today}

\begin{abstract}
We extensively study the localization and the quantum Hall effect
in the Hofstadter butterfly, which emerges in a two-dimensional electron system
with a weak two-dimensional periodic potential.
We numerically calculate the Hall conductivity and the localization length
for finite systems with the disorder in general magnetic fields,
and estimate the energies of the extended levels in an infinite system.
We obtain the Hall plateau diagram on the whole region of
the Hofstadter butterfly, and propose a theory 
for the evolution of the plateau structure with increasing disorder.
There we show that a subband with the Hall conductivity $n e^2/h$
has $|n|$ separated bunches of extended levels, 
at least for an integer $n \leq 2$.
We also find that the clusters of the subbands with
identical Hall conductivity, which repeatedly appear
in the Hofstadter butterfly, have a similar localization property.
\end{abstract}

\maketitle

%\newpage

%Introduction
\section{Introduction}
A two dimensional (2D) electron system with a 2D periodic potential
is expected to exhibit an intricate energy spectrum 
in a strong magnetic field, which is called the Hofstadter butterfly.
When the magnetic length is of the order of the lattice constant,
the interplay of the Landau quantization and Bragg reflection
yields a fractal-like series of the energy gaps,
which depends sensitively on the number of magnetic flux quanta
per a unit cell, \cite{Hofs}
$$\phi = \frac{Ba^2}{h/e},$$
where $B$ is the amplitude of the constant magnetic field
and $a$ is the lattice constant.
Moreover it is shown that each subband carries a quantized Hall
conductivity, which varies with the energy gaps
in a nontrivial manner. \cite{TKNN}
The recent developments in experimental techniques 
make it possible to fabricate a two-dimensional
superlattice on a 2D electron gas.
Evidence of the fractal energy spectrum
was found in a superlattice patterned on 
GaAs/AlGaAs heterostructures, \cite{Albr,Geis}
where the Hall conductivity changes nonmonotonically
as the Fermi energy transfers from one gap to another.

From the theoretical side,
it is intriguing to consider the magnetotransport 
in this intricate energy spectrum.
Broadening of the density of states \cite{Pfan,Wulf} 
and the conductivity \cite{Pfan} 
were investigated for the Hofstadter butterfly
within the self-consistent Born approximation.
For the localization regime,
we expect that, by analogy with unmodulated 2D systems,
the extended levels appear only at certain energies
in the Hofstadter spectrum, and $\sigma_{xy} (E)$ 
will turn into a series of Hall plateaus separated by 
those energies. 

The evolution of the extended states 
as a function of the disorder
was proposed for several flux states in
the Hofstadter butterfly. \cite{Tan,Yang99,Hats}
A finite-size scaling analysis 
was performed for a 2D system modulated by a weak
periodic potential and it was found that
the modulation does not change the critical exponent
at the center of the Landau level. \cite{Huck,Huck2}
Previously we have numerically calculated the
Hall conductivity in weakly modulated 2D systems
at several fluxes and studied the effect of the
localization on the quantum Hall effect in the
Hofstadter butterfly. \cite{Kosh}
There we determined the energies of the
extended levels from the scaling behavior of the Hall conductivity
in finite systems.

In this paper, following the line of the previous work,
we present an extensive study of 
the quantum Hall effect in a single Landau level
in the presence of the two-dimensional periodic potential.
We numerically calculate the Hall conductivity 
and the localization length in general magnetic fluxes,
using the exact diagonalization of the Hamiltonian for finite systems.
We determine the energies of the extended levels in an infinite system
by analyzing the size dependence in the Hall conductivity,
and find that those energies always coincide with
the points where $\sigma_{xy}/(-e^2/h)$ 
becomes $n+1/2$ ($n$ is an integer)
in finite systems.
In particular, we resolved {\it two} separated extended levels
in a subband carrying the Hall conductivity by $2e^2/h$.
We propose a possible scenario 
for the evolution of the extended levels with increasing disorder
for several fluxes, 
which suggests that the Hall plateau structure changes
through the pair annihilation and/or
{\it pair creation} of the extended levels.
We survey the Hall conductivity 
over the whole region of the Hofstadter butterfly,
to find that the electronic structure 
with similar Hall conductivity and localization length
appears in a self-similar manner at different fluxes,
reflecting the fractal property of the ideal spectrum.

% Magnetic Bloch bands
\section{Formulation}
\label{sec_bloch}
We first prepare the formulation to describe a
Bloch electron in magnetic fields.
Let us consider a two-dimensional system in a uniform magnetic field
with a periodic potential $V_p$ and a disorder potential $V_d$,
\begin{equation}
H = \frac{1}{2m}(\Vec{p} + e\Vec{A})^2 + V_p + V_d .
\label{eq_H}
\end{equation}
%The band structure is characterized by a parameter $\phi=Ba^2/(h/e)$, 
%a number of magnetic flux quanta penetrating unit cell \cite{Hofs}.
We assume that $V_p$ has a square form
$$
%\begin{equation}
V_p = V \cos\frac{2\pi}{a}x \!+\! V\cos\frac{2\pi}{a}y,
$$
%\label{vp}
%\end{equation}
and the disorder potential is composed of
randomly distributed $\delta$ potentials $\pm \upsilon_0$
with the number per unit area $n_i$, 
where the amounts of the positive and negative scatterers 
are taken to be equal. 
We consider only the lowest Landau level ($N=0$), assuming that 
the magnetic field is strong enough
and the mixing of the Landau levels is neglected.
In the Landau gauge, the basis can be taken as
\begin{equation}
 |0, k_y \rangle = \sqrt{\frac{1}{\sqrt{\pi}l L_y}}
 \,\,\, \exp(i k_y y) \exp\left[-\frac{(x+k_y l^2)^2}{2l^2}\right],
\end{equation}
with the magnetic length $l = \sqrt{\hbar/eB}$.
The matrix elements of $V_p$ are then written
\begin{eqnarray}
 \langle 0, k'_y | V_p  | 0, k_y \rangle &=& 
\delta_{k'_y,k_y} 2 V e^{-\frac{\pi}{2\phi}}
\cos\frac{k_y a}{\phi} \nonumber\\
&& \hspace{-10mm} + (\delta_{k'_y,k_y+2\pi/a} + \delta_{k'_y,k_y-2\pi/a}) 
V e^{-\frac{\pi}{2\phi}}.
\end{eqnarray}

In an ideal system ($V_d=0$), the wave function can then be expanded as
\begin{equation}
 \psi_{k_y}(\Vec{r}) = \sum_{m} c_m(k_y) | 0, k_y  - \frac{2\pi}{a}m \rangle,
\label{eq_psi}
\end{equation}
with Bloch wave number $k_y$ ranging from $-\pi/a$ to $\pi/a$.
The Schr\"odinger equation is then reduced to Harper's equation,
\begin{eqnarray}
&& V e^{-\pi/2\phi}(c_{m+1} + c_{m-1}) \nonumber \\
&& + 2 V e^{-\pi/2\phi}
\cos\left(\frac{k_y a - 2\pi m}{\phi}\right) c_m  = E c_m.
\label{eq_Harper}
\end{eqnarray}
In a rational flux $\phi = p/q$ ($p,q$ are coprime integers), 
the equation becomes periodic in $m$ with a period $p$,
which corresponds to a distance $qa$ 
in the center coordinate along the $x$ axis,
so that we have the Bloch condition
$c_{m+p} = e^{ik_x qa} c_m$
with another Bloch wave number $k_x$
from $-\pi/(qa)$ to $\pi/(qa)$.
As a result, we have $p$ independent states 
for each of $\Vec{k} = (k_x,k_y)$ in the Landau level,
so that we can label the wave function
as $\psi_{nk}$ on a $q$-folded 
Brillouin zone with the subband index $n = 1,2,...,p$.
We can decompose the wave function as
$\psi_{nk}(\Vec{r}) = e^{i (k_x x + k_y y)} \,\,u_{nk}(\Vec{r})$,
where $u$ satisfies the magnetic Bloch condition
\begin{eqnarray}
 u(x,y+b) &=& u(x,y) \nonumber \\
 u(x+qa,y) &=&  e^{2\pi i p y/a} u(x,y).
\end{eqnarray}

\begin{figure}
\begin{center}
%\leavevmode\includegraphics[width=80mm]{btfl_noimp_ntr.eps}
\leavevmode\includegraphics[width=80mm]{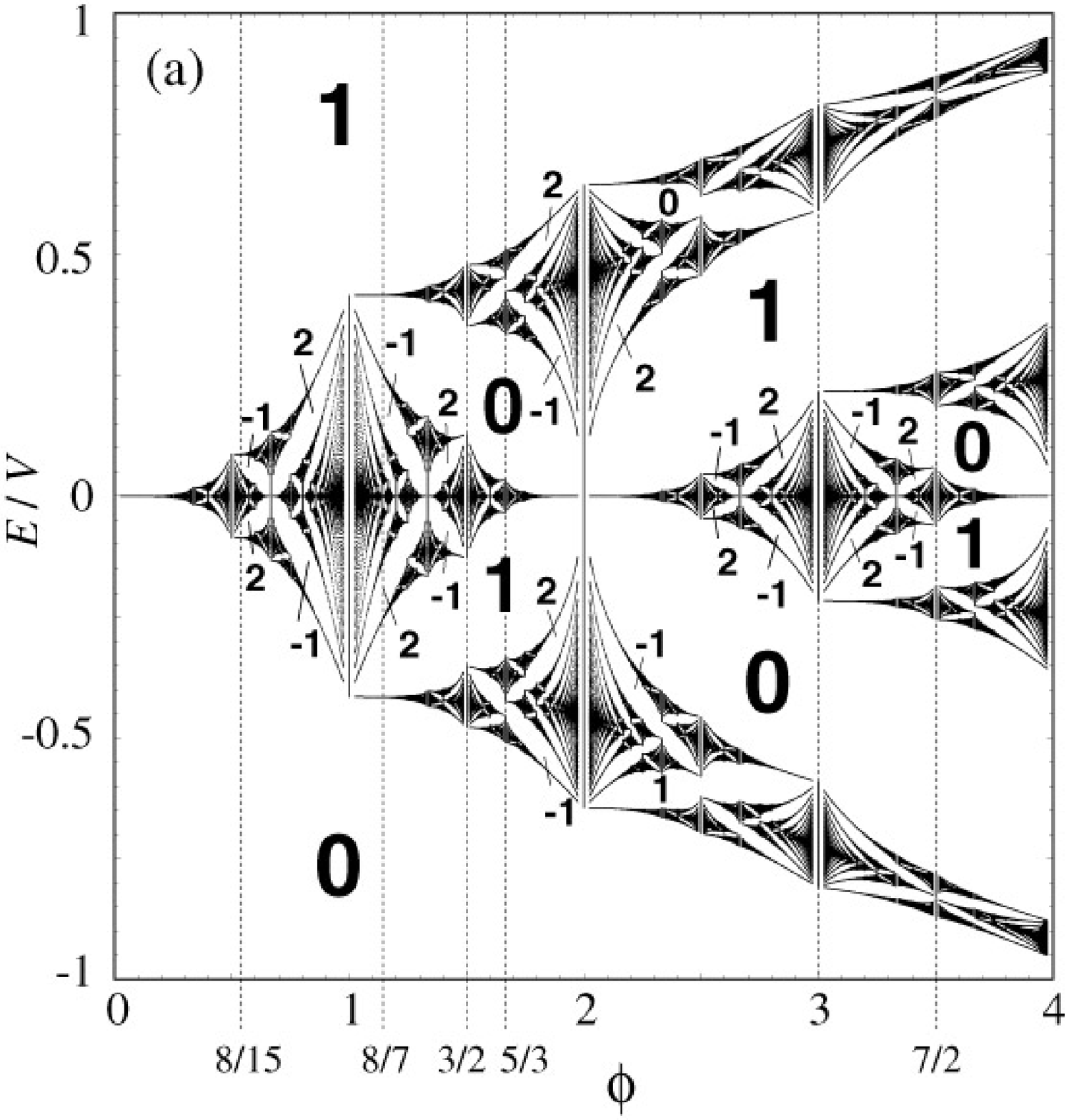}
 \vspace{5mm}

%\leavevmode\includegraphics[width=81mm]{btfl_noimp_wide.eps}
\leavevmode\includegraphics[width=81mm]{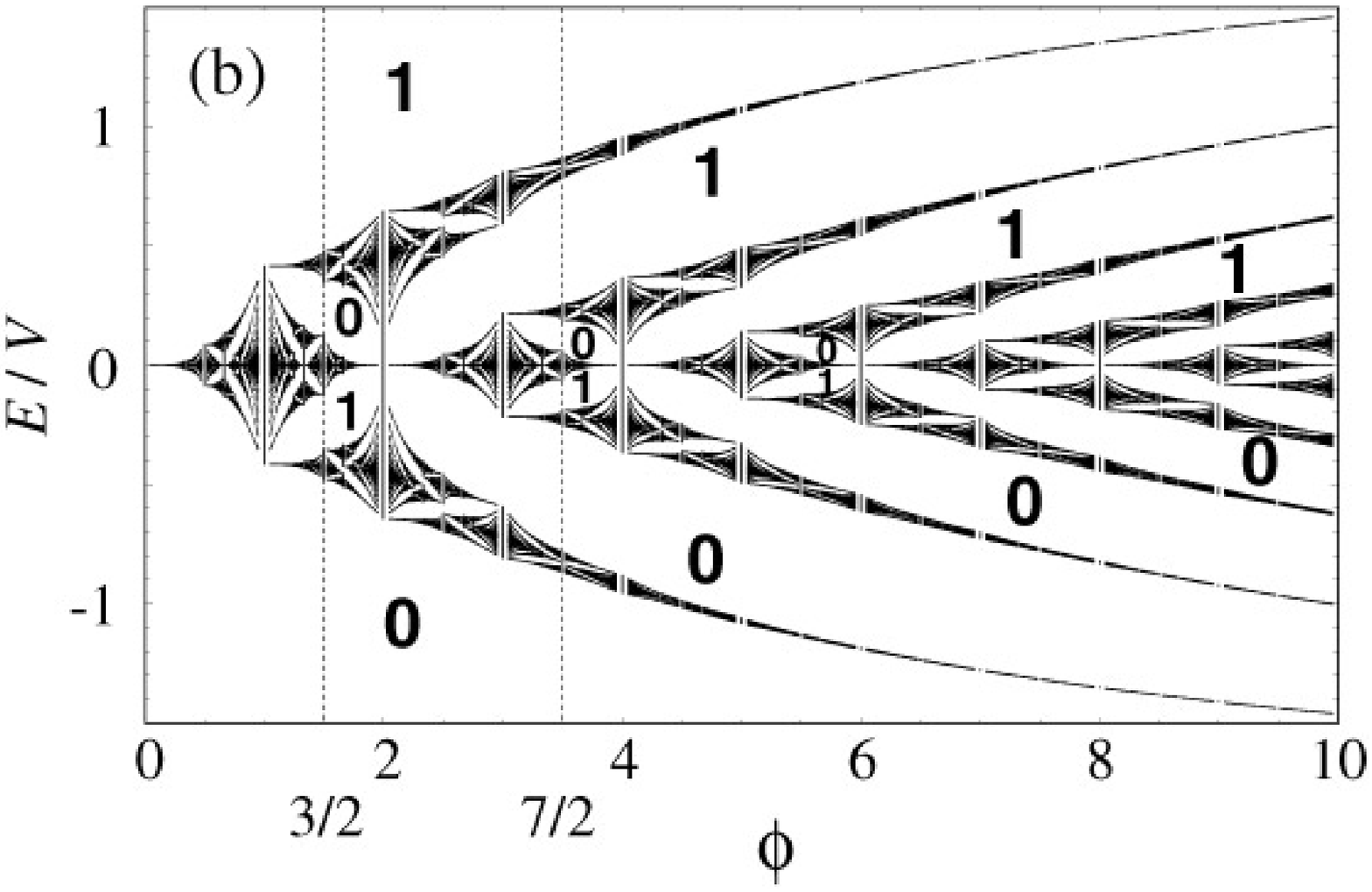}
\end{center}
\caption{(a) Ideal energy spectrum of the lowest Landau level
in a weakly modulated two-dimensional system,
plotted against the magnetic flux $\phi$.
Integers inside gaps indicate the quantized Hall conductivity
in units of $(-e^2/h)$.
(b) Zoom out of the figure (a).
}
\label{fig_btfl}
\end{figure}

Figure \ref{fig_btfl} shows the energy spectrum in the lowest Landau level
at $V_d=0$ plotted against $\phi$.
The intricate band structure is due to 
the number of subbands $p$, which is not a continuous function of $\phi$.
We can show that $p$ subbands never overlap so that we always have
$p-1$ energy gaps inside the Landau level. 
The total width of the spectrum scales with 
the factor $e^{-\pi/2\phi}$ in Eq. (\ref{eq_Harper}),
and shrinks as the flux becomes smaller.
The lower panel shows the zoom out of the spectrum
covering from $\phi=0$ to 10.
We can see that, on going to a higher field,
a series of subbands splits away from 
the center toward higher and lower energies.
These are identified in the semiclassical picture
as the quantization of the electron motion along the equipotential line
around the bottom or the top of the periodic potential.
The widths of those levels become narrower for larger $\phi$
because the coupling between different potential valleys
becomes exponentially small as the magnetic length becomes smaller.

The Hall conductivity $\sigma_{xy}$ is calculated by 
the Kubo formula as
\begin{equation}
 \sigma_{xy} = 
%-\frac{ne}{B} + \frac{\hbar e^2}{iL^2} \!\!\!
\sum_{\epsilon_{\alpha} < E_F}\sum_{\epsilon_{\beta} \neq \epsilon_{\alpha}}
\!\!\! \frac{\langle \alpha | v_x | \beta \rangle 
\langle \beta | v_y | \alpha \rangle 
\!-\! \langle \alpha | v_y | \beta \rangle 
\langle \beta | v_x | \alpha \rangle}
{(\epsilon_{\alpha} - \epsilon_{\beta})^2},
\label{Kubo}
\end{equation}
where $E_F$ is the Fermi energy and
$\epsilon_{\alpha}$ the energy of the eigenstate $|\alpha\rangle$
in the lowest Landau level.
This is rewritten in an ideal system as \cite{TKNN}
\begin{equation}
 \sigma_{xy} = \frac{e^2}{2\pi h} \sum_n \int_{E < E_F} d^2 k 
\,\,\, 2\, {\rm Im}
\left[
\Bigg\langle \partd{u_{nk}}{k_x} \Bigg|
\partd{u_{nk}}{k_y}\Bigg\rangle 
\right],
\label{Kubo_ideal}
\end{equation}
where the summation is taken over all the occupied states.
It is shown that the contribution from all the states in {\it one} subband 
is always quantized in units of $-e^2/h$ \cite{TKNN}.
So we have an integer Hall conductivity
when the Fermi energy $E_F$ is in every gap,
since it simply becomes the summation of the integers for all the subbands
below $E_F$. 
In such a gapped situation, we have a useful expression
for the Hall conductivity called the Str\v{e}da formula, \cite{Stre}
which is directly derived from the Kubo formula (\ref{Kubo}).
This is written as
\begin{equation}
 \sigma_{xy} = -e \frac{\partial n}{\partial B},
\label{eq_Stre}
\end{equation}
where $n$ is the number of states below the Fermi energy
per unit area and $B$ is the magnetic field.
In Fig. \ref{fig_btfl},
we put the value of the Hall conductivity for each of the gaps
obtained by this formula.
The contribution by one subband, $\Delta\sigma_{xy}$,
is obtained as the difference 
in $\sigma_{xy}$ between the gaps above and below the subband.
In Fig. \ref{fig_btfl} (b), the series of branches mentioned above
always carries $\Delta\sigma_{xy} = 0$,
since those subbands, coming from the localized orbitals around
the potential minima or maxima, 
contain a constant number of states $n = 1/a^2$ which is independent of 
the magnetic field.

% Kubo formula (strongly disordered system)
% Self-similar spectrum, phase diagram
\section{Localization and the quantum Hall effect}
\label{sec_qhe}
We now move on to the disordered system to investigate the localization 
and the quantum Hall effect.
Here we numerically diagonalize the Hamiltonian (\ref{eq_H})
of finite systems of $L\times L$ with $L = Ma$ ($M$ is an integer), 
and calculate the localization length $L_{\rm loc}$
and the Hall conductivity  $\sigma_{xy}$
as functions of Fermi energy.
The localization length is obtained from the system size
dependence of the Thouless number $g(L)$,
by assuming that $g(L)$ behaves as $\exp(-L/L_{\rm loc})$,
where $g(L)$ is defined as the ratio of 
the difference in the eigenenergies between periodic and
antiperiodic boundary conditions to the mean level spacing $\Delta E$.
The Hall conductivity is calculated by the Kubo formula
(\ref{Kubo}) with the exact eigenstates in the disordered system,
where the mixing between different Landau levels
is taken into account within the lowest order
in $(V_p + V_d)/(\hbar\omega_c)$ with $\omega_c = eB/m$
being the cyclotron frequency.
Every quantity is averaged over a number of different samples.

In the strong disorder regime,
the energy scale of the disorder is characterized by 
\[
\Gamma = \sqrt{\frac{4n_i v_0^2}{2\pi l^2}},  
\]
which represents the increasing width 
of the Landau level 
in absence of the periodic potential. \cite{Ando74}
This parameter is relevant as long as the disorder 
is strong enough to destroy 
the band structure inside the Landau level.
In the weak disorder limit, on the other hand, 
the relevant parameter is
the increasing width of the Bloch subband,
which is written as
\begin{equation}
 \delta E  \approx 2\pi n_i v_0^2 \rho,
\end{equation}
with $\rho$ being the density of states per unit area.
For a subband in the flux $\phi=p/q$, $\rho$ is approximated as
$1/(Wqa^2)$ with the subband width $W$,
so that we obtain
\begin{equation}
 \delta E \approx \frac{2\pi n_i v_0^2}{W q a^2} \equiv \gamma.
\label{eq_gamma}
\end{equation}
The relation between $\Gamma$ and $\gamma$
is given by
\[
 \gamma = \frac{\pi}{2}\frac{\Gamma^2}{W p}.
\]

% phi=3/2

\begin{figure}
\begin{center}
%\leavevmode\includegraphics[width=80mm]{fig_f32.eps}
\leavevmode\includegraphics[width=80mm]{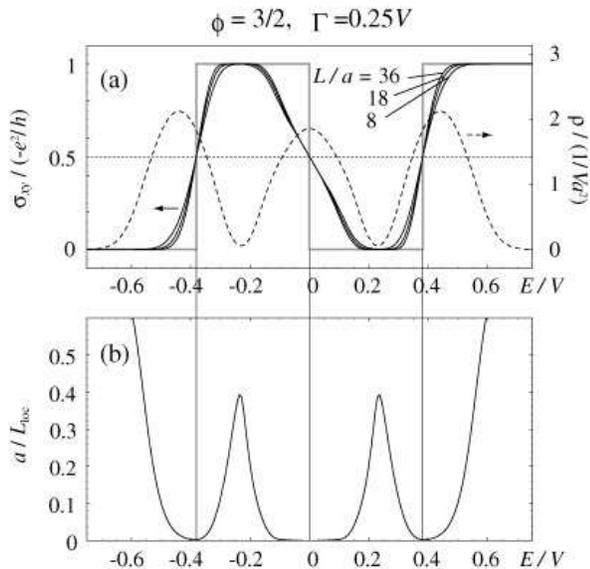}
\end{center}
\caption{
(a) Hall conductivity $\sigma_{xy}$ (solid line)
and the density of states $\rho$ (dashed)
in the flux $\phi = 3/2$ and the disorder $\Gamma/V=0.25$ 
with various sizes $L/a=8$, 18, and 36.
The gray steplike line is an estimate of $\sigma_{xy}$
as $L\rightarrow \infty$.
(b) Inverse localization length
estimated from the Thouless number. 
The vertical lines penetrating the panels 
represent the energies of the extended levels.
}
\label{fig_f32}
\end{figure}

As a typical result, we show the calculation for the
disordered system in the flux $\phi = 3/2$ 
in Fig. \ref{fig_f32}.
The top panel indicates the system size dependence of 
$\sigma_{xy}$ and the density of states, and 
the bottom the inverse localization length $1/L_{\rm loc}$.
The parameter of the disorder is set to $\Gamma/V = 0.25$.
In the clean limit, a Landau level splits into three separated subbands 
since $p=3$, and the Hall conductivity carried by each of the subbands 
is $\Delta\sigma_{xy} = 1,-1,1$.
Here and in the following we show
$\sigma_{xy}$ and $\Delta\sigma_{xy}$ in units of $-e^2/h$.
When the system is disordered 
we find that the gaps between the subbands
are smeared out
while the Hall conductivity converges to quantized values 
around the density of states (DOS) dips, 
indicating the appearance of the Hall plateau.
If we look at the size dependence of the Hall conductivity
in the whole energy region,
we see that $\sigma_{xy}$ always approaches 1 
in the area $\sigma_{xy} > 1/2$ in increasing the size, 
and 0 in $\sigma_{xy} < 1/2$,
so we expect that in an infinite system
the continuous function $\sigma_{xy}(E)$ 
changes into Hall plateaus connected by steps, 
as shown by the steplike line. 
Therefore we speculate that 
the points of $\sigma_{xy}=1/2$ are identified as 
the extended levels in an infinite system, and 
they indeed agree with the energies where localization length diverges 
as shown in Fig. \ref{fig_f32} (b).

We can see similar tendencies 
in the size dependence of  $\sigma_{xy}$ in other fluxes as well,
where the fixed points are always found at  $\sigma_{xy}=n+1/2$
($n$ is an integer) in all the cases investigated.
While each of the three subbands has one extended level
at a certain energy in $\phi=3/2$, 
the localization generally depends on the Hall conductivity assigned to
subbands. We have shown that a subband carrying zero Hall conductivity
is completely localized in finite disorder strength. \cite{Kosh}

\begin{figure}
\begin{center}
%\leavevmode\includegraphics[width=80mm]{trace_f32.eps}
\leavevmode\includegraphics[width=80mm]{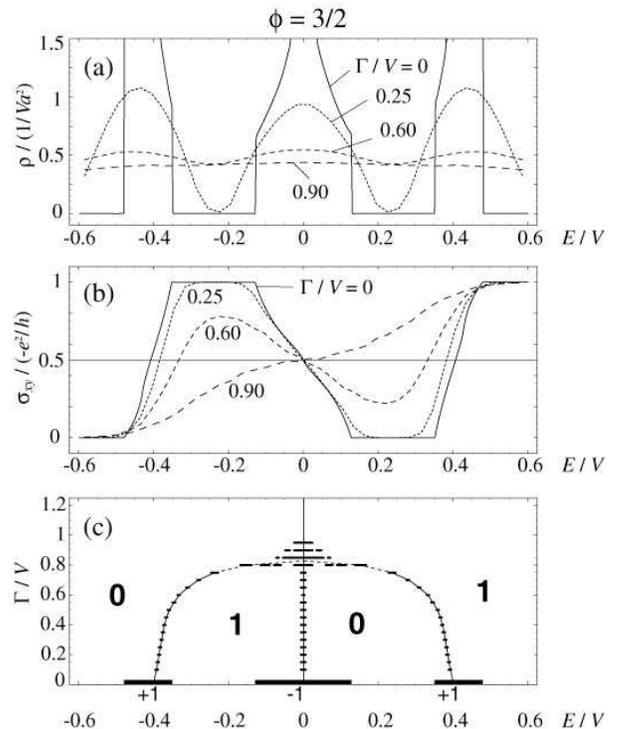}
\end{center}
\caption{
(a) Density of states and (b) the Hall conductivity 
as a function of the Fermi energy
in $\phi = 3/2$ and $L = 12a$
with several disorder parameters $\Gamma/V$.
(c) Trace of the extended levels as a function of disorder,
obtained by taking the energies of $\sigma_{xy} = n+1/2$
(in units of $-e^2/h$).
The horizontal bars show the numerical errors.
Integers in the enclosed areas
represent the quantized Hall conductivities.
The thick bars at $\Gamma =0$ and the integers below them
indicate the ideal subbands and $\Delta\sigma_{xy}$, respectively.
}
\label{trace_f32}
\end{figure}

\begin{figure}
\begin{center}
%\leavevmode\includegraphics[width=80mm]{trace_f53.eps}
\leavevmode\includegraphics[width=80mm]{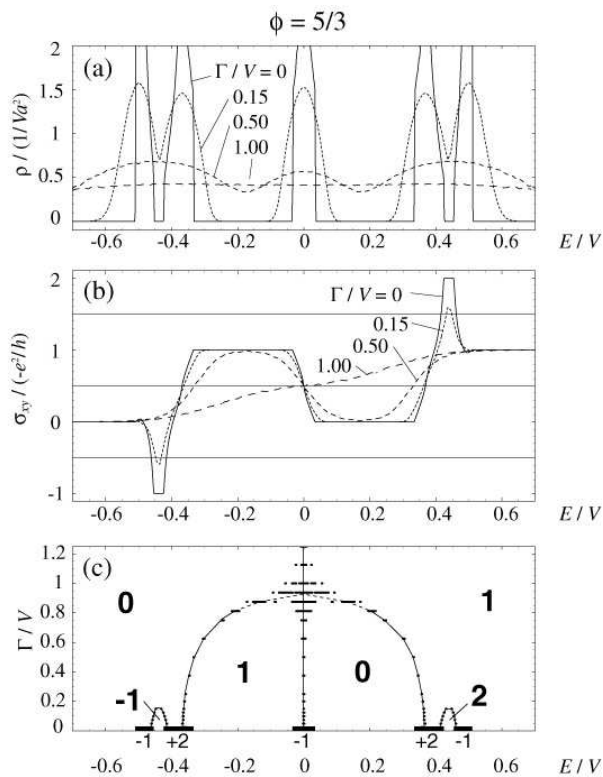}
\end{center}
\caption{Plots similar to Fig. \ref{trace_f32}
calculated for $\phi = 5/3$.
}
\label{trace_f53}
\end{figure}

% Evolution of the extended levels
\section{Evolution of the extended levels}

When we consider a situation where the disorder increases
in the Hofstadter butterfly, we expect that the extended levels
in the subbands move and merge on the energy axis in some way,
and in large enough disorder,
only one remains at the center of the Landau level.
It would be nontrivial and intriguing to ask
how they evolve as a function of disorder,
and how different they are in various fluxes.
Here we study the evolutions for some particular fluxes,
on the assumption that the extended levels always exist
at the energies of $\sigma_{xy} = n+1/2$ with integers $n$,
and the region where $n-1/2 < \sigma_{xy} < n+1/2$ in finite systems
becomes $\sigma_{xy} = n$ in an infinite system.

We first show the results for the flux $\phi=3/2$ in Fig. \ref{trace_f32}.
The top and middle panel indicate
the density of states and the Hall conductivity, respectively,
in a finite system $L = 12a$ for several disorder parameters $\Gamma$.
The bottom panel shows the traces of the extended levels
obtained by just taking the energies of $\sigma_{xy} = n+1/2$.
We observe that three extended levels get closer 
as the disorder increases, and contract into one at a certain $\Gamma$.
The combination of three branches at a time
is due to the electron-hole symmetry
with respect to $E=0$, owing to the equal amounts of the positive and 
negative scatterers. 
If we break the symmetry by introducing imbalance between them,
the evolution changes in such a way that 
two of them annihilate and one is left intact. \cite{Kosh2}

The situation becomes a little complicated in a slightly larger flux 
$\phi = 5/3$, as shown in Fig. \ref{trace_f53}.
In the clean limit, we have 
five subbands carrying $\Delta\sigma_{xy} =-1, +2, -1, +2, -1$ 
from the bottom to the top. Since the second and the fourth bands
pass through $n+1/2$ twice, 
each of them comes to have
two extended levels in sufficiently small disorder on the present assumptions.
On going to stronger disorder, one of the two extended levels
experiences a pair annihilation with the lowest or highest subband,
and only three extended levels are left as in $\phi = 3/2$.
When we start with a more complicated flux, 
we speculate that the Hall plateau structure is simplified 
one by one, going through analogs of the simpler fluxes nearby.

% phi=8/15

\begin{figure}
\begin{center}
%\leavevmode\includegraphics[width=80mm]{fig_f815.eps}
\leavevmode\includegraphics[width=80mm]{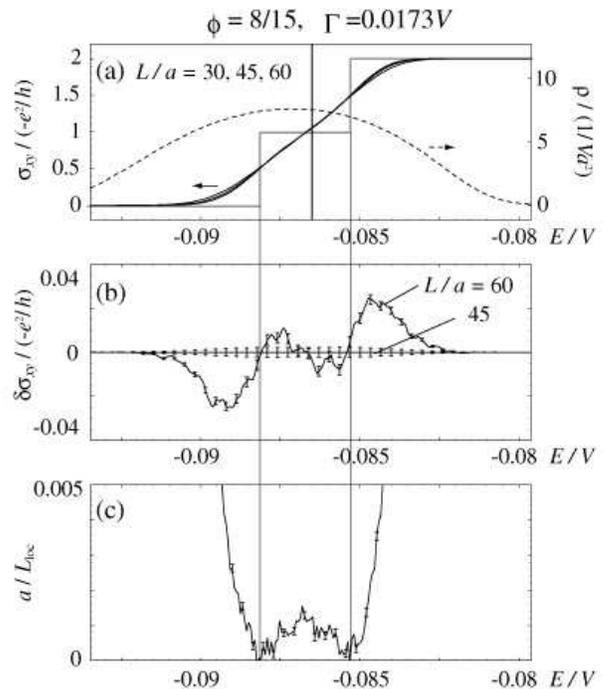}
\end{center}
\caption{
(a) Hall conductivity $\sigma_{xy}$
and the density of states $\rho$
calculated for the lowest subband
in $\phi = 8/15$, carrying $\Delta\sigma_{xy} = -2e^2/h$,
with the disorder $\Gamma/V = 0.0173$
and various sizes $L/a=30$, 45, and 60.
The gray steplike line is an estimate at $L\rightarrow \infty$.
The vertical line at $E\approx -0.0865$
represents the energy region of the ideal subband.
(b) Relative value of $\sigma_{xy}$ in $L/a=60$
estimated from $L/a=45$.
(c) Inverse localization length (in units of $a$)
estimated from the Thouless number. 
Two vertical lines penetrating the panels
represent the energies of the extended levels 
in an infinite system.
}
\label{fig_f815}
\end{figure}

To check that there are actually two separated extended levels
in a subband carrying $\Delta\sigma_{xy}=2$,
we can make a scaling analysis in the Hall conductivity
on varying the system size.
In the flux $\phi = 5/3$, unfortunately, 
the large statistical error prevents us from resolving 
the splitting,
but we can in a similar situation in another flux $\phi= 8/15$,
where the lowest subband carrying $\Delta\sigma_{xy} = 2$.
Figure \ref{fig_f815} shows the Hall conductivity and the
localization length for the lowest subband 
with disorder $\Gamma = 0.0173V$.
The difference in $\sigma_{xy}$ between $L/a = 45$ and 30,
shown in the middle panel, indicates that $\sigma_{xy}$ actually has 
turning points around $\sigma_{xy}=1/2$ and $3/2$. 
This is confirmed by the calculation of the localization length
in the lowest panel, showing that 
$L_{\rm loc}$ diverges at those two energies,
so it is quite likely that there are two extended levels 
in this subband in an infinite system.

A major difference from $\phi=5/3$ is that 
the ideal subband width is much narrower than the energy gap 
in the present case, so we can set the disorder
in such a way that the states in the subband are completely mixed up
but not with other subbands.
We speculate that this situation makes the localization length
smaller and enables us to resolve the separated extended levels
in a finite size calculation.
We have another example where one subband
has more than one extended levels
in an {\it anisotropic} modulated quantum Hall systems.
There $\sigma_{xy}(E)$ in a subband with $\Delta\sigma_{xy} = 1$
has a non-monotonic behavior crossing $\sigma_{xy} = 1/2$ three times,
so that three extended levels arise. \cite{Kosh}

\begin{figure}
\begin{center}
%\leavevmode\includegraphics[width=80mm]{trace_f87.eps}
\leavevmode\includegraphics[width=80mm]{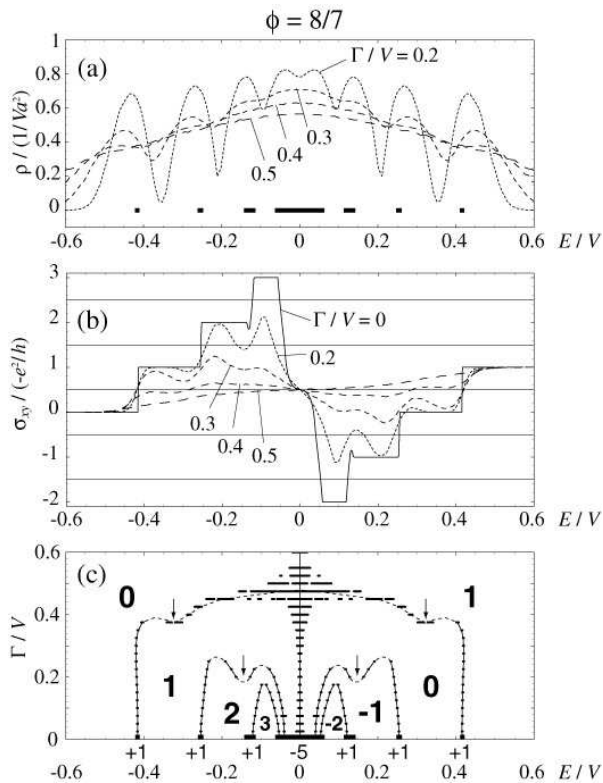}
\end{center}
\caption{
Plots similar to Fig. \ref{trace_f32}
calculated for $\phi = 8/7$.
Arrows indicate the positions of the pair creations
of the extended levels.
}
\label{trace_f87}
\end{figure}

\begin{figure}
\begin{center}
%\leavevmode\includegraphics[width=80mm]{fig_pair.eps}
\leavevmode\includegraphics[width=80mm]{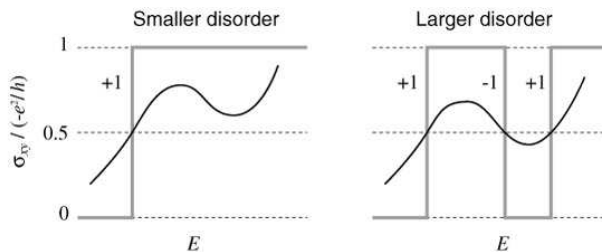}
\end{center}
\caption{Schematic figure of pair creation
of the extended levels.
The left and right panels show $\sigma_{xy}(E)$ 
in smaller and larger disorder, respectively,
where a solid curve is for a finite system 
and a gray line for an infinite system.
The pair of extended levels are newly created
when a dip touches the line of $\sigma_{xy} = 0.5 (-e^2/h)$.}
\label{fig_pair}
\end{figure}

\begin{figure}
\begin{center}
%\leavevmode\includegraphics[width=80mm]{fig_f54.eps}
\leavevmode\includegraphics[width=80mm]{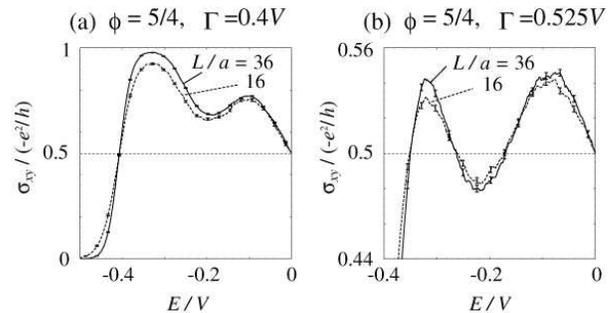}
\end{center}
\caption{Hall conductivity $\sigma_{xy}$
in the flux $\phi = 5/4$ and the disorder
$\Gamma/V$ (a) 0.4 and (b) 0.525
(in different vertical scales),
with the system sizes $L/a=16$ and 36.
}
\label{fig_f54}
\end{figure}

% phi=8/7

Last we investigate a flux $\phi =8/7$, slightly away
from $\phi=1$. Around this region the Landau level
splits into a number of tiny subbands 
as seen in the original Hofstadter butterfly,
just like the Landau levels in the usual 2DEG around zero field. 
This is actually understood 
as the Landau quantization in the magnetic Bloch band 
at $\phi = 1$ caused by the residual flux
$\phi-1$. \cite{Chan}
As seen in Fig. \ref{trace_f87},
each of these ``Landau levels'' equally carries the Hall conductivity
of $\Delta\sigma_{xy} = 1$ 
while the center one has a large negative value.
It is highly nontrivial how the extended states 
evolves in increasing disorder in such a situation.

Figure \ref{trace_f87} shows the disorder dependence of 
the density of states, the Hall conductivity $\sigma_{xy}$,
and the traces of $\sigma_{xy} = n+1/2$ in a fixed system size $L=14a$
in $\phi =8/7$.
Remarkably the result suggests that
a pair creation of extended states can occur 
in increasing disorder, as indicated by the arrows in (c).
In (b), we can see that this happens when a dip in $\sigma_{xy}(E)$  
touches the line of $\sigma_{xy} = n + 1/2$,
where a pair of extended levels with opposite Hall conductivity
are created as schematically shown in Fig. \ref{fig_pair}.
In even stronger disorder, these newly created pairs
annihilate with other extended levels
in the lower and higher subbands.

To confirm the existence of a pair creation,
we investigate the scaling behavior of the Hall conductivity $\sigma_{xy}(E)$
in a similar situation in $\phi = 5/4$,
for the smaller disorder $\Gamma = 0.4 V$ and the larger $0.525V$.
The results in Fig. \ref{fig_f54} show that
the direction of the size dependence around the dip ($E/V \approx -0.2$)
seems to change when passing through $\sigma_{xy} = 0.5$,
within the system size numerically available.
This is consistent with the proposed scenario as in Fig. \ref{fig_pair},
while we need further extensive work to confirm
that the tendency remains towards the infinite system.

For the usual 2DEG without a periodic potential,
there has been a long debate about 
the fate of the extended states in the Landau levels
in the strong disorder limit in a weak magnetic field.
It has been argued that all the extended states 
float up in the energy axis with increasing disorder
and the system finally becomes an insulator. \cite{Levi,Khme,Laug}
Numerical calculations in tight-binding lattice models,
on the other hand, showed that the 
the states with negative Hall conductivity move down
from the tight-binding band center,
to annihilate with the extended states in the 
lower Landau levels. \cite{Ando89,Liu,Shen}
Our theory may suggest a somewhat different story
where pair creations of the extended states occur,
since we actually observe similar dips in $\sigma_{xy}(E)$
around higher Landau levels in a disordered nonperiodic 2DEG.

We also remark that the evolution of the extended levels
generally depends on the correlation lengths of 
the disorder potential.
It was shown in the tight-binding model that
the pair annihilation of the extended states
with positive and negative Hall conductivity
do not occur in the long range disorder. \cite{Kosc,Shen2}
In our problem, the scenario of the pair creation might change 
in a correlating disorder,
since the appearance of the dips in $\sigma_{xy}(E)$ is
due to the inter-Landau-level mixing \cite{Ando75,Ando86}, and thus 
is suppressed by the long-range scatterers.
This should be addressed in future work.

% Phase diagram
\section{Hall plateau diagram}
\label{sec_pla}
\begin{figure}
\begin{center}
%\leavevmode\includegraphics[width=70mm]{fig_phase_g05.eps}
\leavevmode\includegraphics[width=70mm]{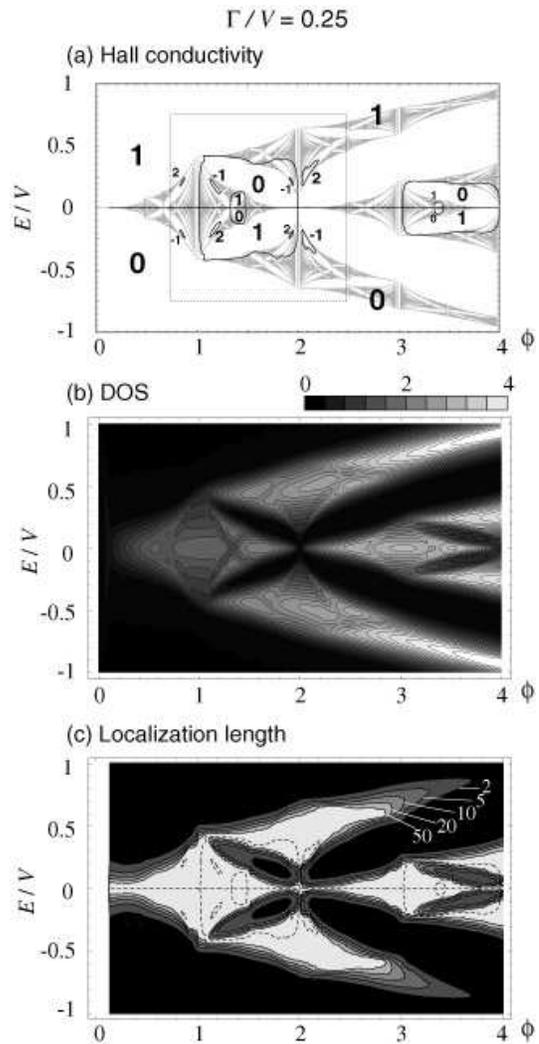}
\end{center}
\caption{
(a) Hall plateau diagram in the lowest Landau level
at the disorder $\Gamma/V=0.25$, plotted against 
the magnetic flux and the Fermi energy.
Solid lines show the energy of $\sigma_{xy} = n+1/2$ ($n$ is an integer),
which are identified as the extended levels in an infinite system,
and the integers indicate the Hall conductivity in units of $-e^2/h$.
The ideal spectrum is shown as the gray scale.
(b) Corresponding plots for the density of states in units of $1/(Va^2)$.
(c) Localization length in units of $a$. 
The dashed lines indicate the extended levels shown in (a).
}
\label{fig_phase_g05}
\end{figure}

We extend the analysis to general fluxes 
to look into the quantum Hall effect
over the whole region of the Hofstadter butterfly.
Figure \ref{fig_phase_g05} shows 
(a) the Hall conductivity,
(b) the density of states, and (c) the localization length
in disordered systems with $\Gamma/V = 0.25$,
plotted against the Fermi energy and the magnetic field.
The solid lines in Fig. \ref{fig_phase_g05} (a) represent the extended levels
in an infinite system, and the enclosed areas are the Hall plateaus
with the quantized value of $\sigma_{xy}$,
where we again assume that 
the extended levels in an infinite system 
coincide with the energies of $\sigma_{xy} = n+1/2$
in a finite system.
In the density of states [Fig. \ref{fig_phase_g05}(b)], 
we recognize valleys in the contour plot
as the remnants of the minigaps,
and some of them can be associated with the Hall plateaus in (a).
For instance, a pair of valleys between $\phi=1$ and 2
inside the Landau level
corresponds to two major plateaus
with $\sigma_{xy} = 0$ in $0<E<0.5V$ and $\sigma_{xy} = 1$ in $-0.5V<E<0$.

In Fig. \ref{fig_phase_g05}(c),
we see that the energy region with large 
localization length becomes broad
particularly around integer fluxes.
There the cluster of subbands widely spreads along the 
energy axis, and thus the states are less easily mixed.
We also find that the upper and lower branches of the spectrum
in $\phi \gtrsim 2$, carrying zero net Hall conductivity,
are all localized, as seen in the particular case $\phi=3$. \cite{Kosh}
The localization length is smaller for larger $\phi$ in those branches,
mainly because the mixing of the states becomes stronger in narrower subbands.

\begin{figure}
\begin{center}
%\leavevmode\includegraphics[width=80mm]{fig_phase_g03-1.eps}
\leavevmode\includegraphics[width=80mm]{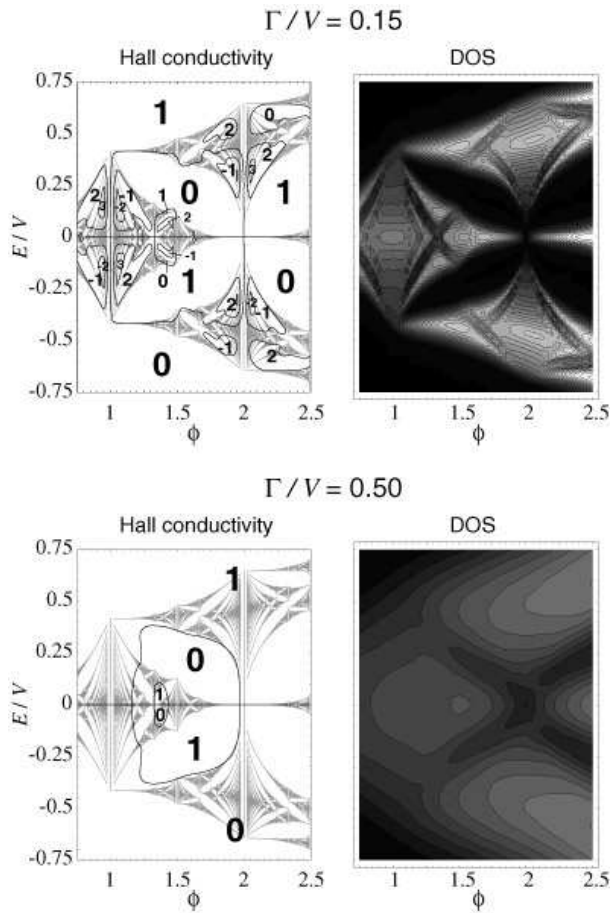}
\end{center}
\caption{
Hall plateau diagram (left)
and the density of states (right)
in the disorder $\Gamma/V=0.15$ (top) and 0.5 (bottom),
plotted against 
the magnetic flux and the Fermi energy.
The corresponding parameter region 
is shown as the dashed line in Fig. {\ref{fig_phase_g05}}(a).
}
\label{fig_phase_g03-1}
\end{figure}

In Fig. \ref{fig_phase_g03-1}, we show the Hall plateau diagrams 
and the density of states
for the different disorder parameters $\Gamma/V = 0.15$ and 0.5,
which exhibit the dependence on $\Gamma$
together with Fig. \ref{fig_phase_g05}(a) and \ref{fig_phase_g05}(b). 
We can see that
the small Hall plateaus coming from the fine gap structure
gradually disappear as the disorder becomes larger, 
and the only extended level is left
at the center of the Landau level in the strong disorder limit.
We here notice in $\Gamma/V = 1.5$ that the two largest plateaus
between $\phi=1$ and 2 mentioned above 
are more easily destroyed around $\phi=1$ than around $\phi=2$,
or the right end is detached from
$\phi=1$ while the left sticks to  $\phi=2$.
This is because the tiny gaps around $\phi =1$ are easily 
swallowed up by the large density of states around the center.

% similarity
\section{Self-similarity}
\label{sec_self}

The interesting observation in the Hofstadter butterfly
is that the identical spectrum
with {\it the identical distribution of the Hall conductivity}
repeatedly appears in different fluxes.
In Fig. \ref{fig_btfl}(b), we can see that 
the spectrum and the Hall conductivity 
at $\phi$ correspond to the middle part of $\phi+2$
with the top and bottom branches excluded,
so that the entire structure repeats over $\phi+2m$ 
with integer $m$.
The proof of the correspondence is presented in Appendix.
We note that, while the Hofstadter butterfly 
has similar gap structures everywhere in a fractal fashion, 
the Hall conductivities in the corresponding gaps
do not always coincide.
An example of the incomplete correspondence is seen 
between $\phi$ and $\phi' = 1/(1-1/\phi)$ (such as $\phi=3$ and $3/2$).
In the tight-binding model, it was shown that
the distributions of the Hall conductivity
within a cluster resemble each other up to a scale factor
among some series of fluxes. \cite{Ritt}

\begin{figure}
\begin{center}
%\leavevmode\includegraphics[width=80mm]{comp_f32-72.eps}
\leavevmode\includegraphics[width=80mm]{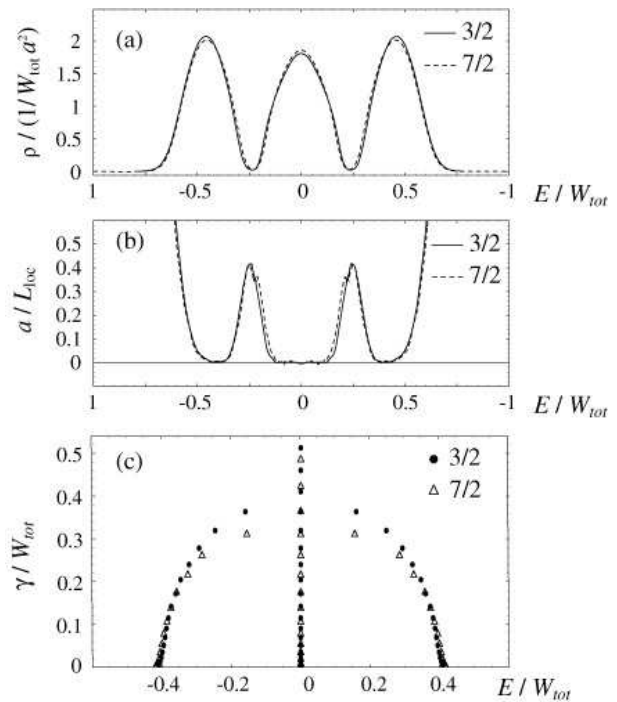}
\end{center}
\caption{
(a) Density of states 
and (b) the inverse localization length
calculated for disordered systems with $\phi = 3/2$ (solid lines)
and 7/2 (dashed).
The energy scale is normalized by the whole width of the Landau level
in the ideal system, $W_{\rm tot}$, and 
the density of states in units of $1/(W_{\rm tot} a^2)$.
The parameter of the disorder is $\gamma/W_{\rm tot} = 0.035$.
(c) Evolutions of the extended levels with changing disorder
$\gamma$, in $\phi=3/2$ (filled circles) and 7/2 (triangles).  
}
\label{fig_comp}
\end{figure}

In the following we show that, in the presence of disorder,
the corresponding clusters with identical Hall conductivity
have qualitative agreement also in the localization length.
Here we particularly take a pair of fluxes $\phi=3/2$ and $7/2$, 
in both of which the central three subbands have
the Hall conductivity $(1,-1,1)$.
In Fig. \ref{fig_comp} we compare the disorder effects
on the density of states and the localization length.
We set the disorder as $\Gamma = 0.25$ and 0.20
for $\phi=3/2$ and $7/2$, respectively,
so that the renormalized DOS broadening $\gamma/W_{\rm tot}$
is equivalent, where we defined $\gamma$ by putting in Eq. (\ref{eq_gamma})
the full width of three subbands $W_{\rm tot}$.
The result shows that the densities of states
are broadened equivalently as expected,
and that the localization lengths $L_{\rm loc}$ 
then agree qualitatively without any scale factors.

Figure \ref{fig_comp}(c) shows the evolution of the extended levels
as a function of the disorder,
which are obtained by taking the points of $\sigma_{xy} = 1/2$.
Three extended levels in $\phi = 3/2$ and $7/2$ 
evolve in a parallel fashion with the disorder strength $\gamma/W_{\rm tot}$,
where they come closer as $\gamma$ becomes larger
and combine into one at $\gamma/W_{\rm tot} \approx 0.4$.
The critical disorder at which the combination occurs
is slightly smaller in $\phi=7/2$ than in $3/2$,
presumably because
in $\phi=7/2$ the level repulsion from the outer subbands (out of the figure)
pushes the states in the central three subbands
toward the center of the spectrum and that enhances 
the contraction of the extended levels.
We expect that the critical behavior of the three extended levels
at the combing point is universal, but we could not estimate
the critical exponent in this simulation
due to statistical errors.
The evolution of the plateau diagram should 
become basically similar among $\phi + 2m$, so 
we know all from the information of the first unit.

\begin{figure}
\begin{center}
%\leavevmode\includegraphics[width=70mm]{btfl_schem.eps}
\leavevmode\includegraphics[width=70mm]{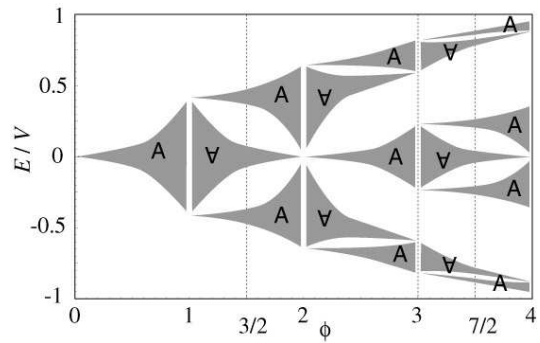}
\end{center}
\caption{Schematic diagram showing 
decomposition of the Hofstadter butterfly
into identical units, where the gap structure and 
the Hall conductivity inside each gap coincide.
The letter A indicates the direction.
}
\label{fig_schem}
\end{figure}

The similar subband structures
with the identical Hall conductivity can be found 
in other hierarchies in the Hofstadter butterfly.
The smallest unit that we have found is schematically 
shown in Fig. \ref{fig_schem},
where the structure indicated by A between $\phi=0$ and 1
repeats over the entire region.
We can show similarly that each unit has the identical 
distribution of the number of states to each subband
so that we have the identical Hall conductivity for every gap.
We expect that the localization length in each single unit A
becomes similar when the disorder is weak enough
that the mixing among different units can be neglected.

% Conclusion
\section{Conclusion}
We studied the quantum Hall effect in a Landau level
in the presence of a two-dimensional periodic potential
with short-range disorder potentials.
It is found that, in all the cases we studied, 
the Hall conductivity becomes size independent 
at $\sigma_{xy} = n+1/2$ (in units of $- e^2/h$), 
and those points are identified 
as the extended levels in an infinite system.
We propose a possible model for the
evolution of the extended levels by tracing $\sigma_{xy} = n+1/2$,
which predicts that a subband with $\sigma_{xy} = n$
has $n$ (or more) bunches of extended levels,
and possibly that pair creation of the extended levels can occur
in increasing disorder, as well as pair annihilation.
We also find that the clusters of subbands
with an identical Hall conductivity, which
compose the Hofstadter butterfly in a fractal fashion,
have a similar localization length
in the presence of the disorder.

% Acknowledgments
\section*{ACKNOWLEDGMENTS}
%%%%%%%%%%%%%%%%%%%%%%%%%%%%%%%%%%%%%%%%%%%%%%%%%%%%%%%%%%%%%%%%%%%%%%%%%%%%%%%
%
%%%%%%%%%%%%%%%%%%%%%%%%%%%%%%%%%%%%%%%%%%%%%%%%%%%%%%%%%%%%%%%%%%%%%%%%%%%%%%%
This work has been supported in part by the 21st Century COE Program at
Tokyo Tech \lq\lq Nanometer-Scale Quantum Physics'' and by a Grant-in-Aid
for COE (12CE2004 \lq\lq Control of Electrons by Quantum Dot Structures
and Its Application to Advanced Electronics'') from the Ministry of
Education, Science and Culture, Japan.  
One of the authors (M.K.) is supported by a Grant-in-Aid for 
Scientific Reseach from 
the Ministry of Education, Science and Culture, Japan.
Numerical calculations were
performed in part using the facilities of the Supercomputer Center,
Institute for Solid State Physics, University of Tokyo.  \par
%%%%%%%%%%%%%%%%%%%%%%%%%%%%%%%%%%%%%%%%%%%%%%%%%%%%%%%%%%%%%%%%%%%%%%%%%%%%%%%
%
%%%%%%%%%%%%%%%%%%%%%%%%%%%%%%%%%%%%%%%%%%%%%%%%%%%%%%%%%%%%%%%%%%%%%%%%%%%%%%%

\appendix*
\section{Concidence in the Hall conductivity}
\label{sec_app}
The equivalence in the gap structure 
with the Hall conductivity between
$\phi$ and $\phi+2$, discussed in Sec. \ref{sec_self},
is explained as follows.
The number of subbands in a Landau level
is given by the numerator of the magnetic flux $\phi$,
so that we have $p$ bands in $\phi=p/q$ and $p+2q$ bands in $\phi+2$.
Each single subband in $\phi$ and $\phi+2$ consists of
the equal number of states per unit area, $1/(qa^2)$,
since two fluxes have a common denominator $q$
so that they have equal foldings of the Brillouin zone.
We can then see that each of the top and bottom branches in $\phi+2$
contains $q$ subbands,
because each has a constant number of states $1/a^2$ as explained in Sec. II.
Thus the number of subbands in the middle part in $\phi+2$
(the top and bottom branches removed) becomes $(p+2q)-2\times q = p$,
which is equal to the total subbands in $\phi$.
Now we see that the corresponding spectra between $\phi$ and $\phi+2$
have the same number of subbands with the the equal number of states,
so we come to the conclusion that the Hall conductivity 
becomes identical between the corresponding gaps,
by using the Str\v{e}da formula (\ref{eq_Stre}).


\begin{thebibliography}{99}
\bibitem{Hofs}
D. R. Hofstadter, Phys. Rev. B {\bf 14}, 2239 (1976).
\bibitem{TKNN}
D. J. Thouless, M. Kohmoto, M. P. Nightingale, and M. den Nijs,
Phys. Rev. Lett. {\bf 49}, 405 (1982).
\bibitem{Albr}
C. Albrecht, J. H. Smet, K. von Klitzing, D. Weiss, V. Umansky,
and H. Schweizer, Phys. Rev. Lett. {\bf 86}, 147 (2001);
Physica E {\bf 20}, 143 (2003).
\bibitem{Geis}
M. C. Geisler, J. H. Smet, V. Umansky, K. von Klitzing,
B. Naundorf, R. Ketzmerick, and H. Schweizer,
Phys. Rev. Lett. {\bf 92}, 256801 (2004).
\bibitem{Pfan}
D. Pfannkuche and R. R. Gerhardts, Phys. Rev. B {\bf 46},
12606 (1992).
\bibitem{Wulf}
U. Wulf and A. H. MacDonald, Phys. Rev. B {\bf 47},
6566 (1993).
\bibitem{Tan}
Y. Tan, J. Phys. Condens. Matter {\bf 6}, 7941 (1994);
Phys. Rev. B {\bf 49}, 1827 (1994).
\bibitem{Hats}
Y. Hatsugai, K. Ishibashi, and Y. Morita, 
Phys. Rev. Lett. {\bf 83}, 2246 (1999).
\bibitem{Yang99}
K. Yang and R.N. Bhatt, Phys. Rev. B {\bf 59}, 8144 (1999). 
\bibitem{Huck} 
B. Huckestein,  Phys. Rev. Lett. {\bf 72}, 1080 (1994).
\bibitem{Huck2} 
B. Huckestein,  Rev. Mod. Phys. {\bf 67}, 357 (1995).
\bibitem{Kosh}
M. Koshino and T. Ando, J. Phys. Soc. Jpn. {\bf 73}, (2004) 3243.
\bibitem{Stre}
P. St\v{r}eda, J. Phys. C {\bf 15}, L718 (1982).
\bibitem{Ando74}
T. Ando and Y. Uemura, J. Phys. Soc. Jpn. {\bf 36}, 959 (1974).
\bibitem{Kosh2}
M. Koshino and T. Ando, Physica E, {\bf 29}, 588 (2005).
\bibitem{Chan}
M. Chang and Q. Niu, Phys. Rev. Lett. {\bf 75}, 1348 (1995).
\bibitem{Levi}
H. Levine, S. B Libby, and A. M. M. Pruisken, 
Phys. Rev. Lett. {\bf 51}, 1915 (1983).
\bibitem{Khme}
D. E. Khmelnitskii, Phys. Lett. {\bf 106A}, 182 (1984). 
\bibitem{Laug}
R. B. Laughlin, Phys. Rev. Lett. {\bf 52}, 2304 (1984). 
\bibitem{Ando89}
T. Ando, Phys. Rev. B {\bf 40}, 5325 (1989). 
\bibitem{Liu}
D. Z. Liu, X. C. Xie, and Q. Niu, Phys. Rev. Lett. {\bf 76}, 975 (1996).
\bibitem{Shen}
D. N. Sheng and Z. Y. Weng, Phys. Rev. Lett. {\bf 78}, 318 (1997). 
\bibitem{Shen2}
D. N. Sheng, Z. Y. Weng, and X. G. Wen, Phys. Rev. B {\bf 64}, 165317 (2001). 
\bibitem{Kosc}
T. Koschny, H. Potempa, and L. Schweitzer, Phys. Rev. Lett. {\bf 86}, 3863 (2001).
\bibitem{Ando75}
T. Ando, Y. Matsumoto, and Y. Uemura, 
J. Phys. Soc. Jpn. {\bf 39}, 279 (1975).
\bibitem{Ando86}
T. Ando, J. Phys. Soc. Jpn. {\bf 55}, 3199 (1986).
\bibitem{Ritt}
H. Ritter, Z. Phys. B {\bf 56}, 185 (1984).

\end{thebibliography}
\end{document}